\documentstyle[11pt,aaspp4,flushrt]{article}

\newcommand{\andre}{Andr\'{e}}
\newcommand{\lehar}{Leh\'{a}r}

\begin{document}

\title{Ring-Like Structure in the Radio Lobe of MG0248+0641}

\centerline{\it Submitted to AJ: August 15, 1997; Accepted: September 30, 1997}

\author{Samuel R. Conner\altaffilmark{1}, Asantha R. Cooray\altaffilmark{1,2}, \andre\, B. Fletcher\altaffilmark{1}, Bernard F. Burke\altaffilmark{1}, Joseph \lehar\,\altaffilmark{3}, Peter M. Garnavich\altaffilmark{3,4}, Tom W. B. Muxlow\altaffilmark{5},
 Peter Thomasson\altaffilmark{5}, John P. Blakeslee\altaffilmark{6}}

\altaffiltext{1}{Department of Physics,
Massachusetts Institute of Technology, 77 Massachusetts Avenue,
Cambridge MA 02139, USA.}
\altaffiltext{2}{Now at Department of Astronomy and Astrophysics,
University of Chicago, Chicago IL 60637, USA. E-mail: asante@hyde.uchicago.edu.}
\altaffiltext{3}{Center for Astrophysics, 60 Garden Street, Cambridge MA 02138, USA.}
\altaffiltext{4}{Visiting Astronomer, Kitt Peak National Observatory, National Optical
Astronomical Observatories, which is operated by the Association of Universities for
Research in Astronomy (AURA) under cooperative agreement with the
National Science Foundation.}
\altaffiltext{5}{Nuffield Radio Astronomy Laboratories, Jodrell Bank, Cheshire SK11 9DL, UK.}
\altaffiltext{6}{Palomar Observatory, California Institute of Technology, MS 105-24, Pasadena CA 91125, USA.}

\begin{abstract}

We present radio and optical observations of MG0248+0641, which contains
a kiloparsec-scale ring-like structure in one of its radio lobes.
The radio observations show a typical core-double
morphology: a central core between two lobes, each of which has a
hotspot. The western radio lobe
appears as a nearly continuous
ring, with linear polarization electric field
vectors which are oriented in a radial
direction from the ring center.
We consider several different interpretations for the nature
of this ring,  including gravitational lensing of
a normal jet by a foreground galaxy.
Even though simple lensing models can describe the ring
morphology reasonably well, the high linear polarization seen around the ring
cannot be easily explained, and no lensing object has yet been found
in deep optical and infrared searches within the extent of the ring.
If the radio ring is indeed caused by
gravitational lensing, the implied mass-to-light ratio is typical of the very
high values seen in other candidate ``dark'' gravitational lenses.
The chance interposition of a galactic supernova
remnant, nova, planetary nebula, or H II region, has been ruled out.
The highly polarized ring of MG0248+0641 is much like the prominent ring 
seen in 3C219, and the multiple ones in 3C310 and Hercules A, suggesting 
that similar
physical processes are producing shell structures in these radio galaxies.
The ring in MG0248+0641 may be caused by the formation of ``bubbles'',
as a result of instabilities in the energy flow down the western
radio jet.
It may also be possible that the required
instabilities are triggered by the infall of
gas, via tidal interaction of the central source  with a nearby galaxy.
This scenario may be indicated by our marginal detection of an optical
source close to the western hotspot.

\end{abstract}

\keywords{ radio galaxies: individual (MG0248+0641)
           --- active galactic nuclei - radio jets and lobes.
           }

\section{Introduction}

The MIT-Green Bank-VLA (MG-VLA)
lens search surveys
(Lawrence {\it et al.} 1986; Hewitt 1986; \lehar\, 1991;
Herold-Jacobson 1996)
 have so far
 produced six confirmed lenses from snapshot maps of about 6000 sources:
 MG0414+0534 (Hewitt {\it et al.} 1992),
MG0751+2716 (\lehar\, {\it et al.} 1997),
MG1131+0456 (Hewitt {\it et al.} 1988),
MG1549+3047 (\lehar\, {\it et al.} 1993),
MG1654+1346 (Langston {\it et al.} 1989), and
MG2016+112 (Lawrence {\it et al.} 1984).
After an extensive improvement in mapping procedures
the same data in the MG-VLA survey have
been reanalyzed, the result being a promising subsample of
radio sources with morphologies typical of gravitational lensing (Conner {\it
et al.} 1993).
Included in this was MG0248+0641, which, after recalibration and
remapping of the original
6 cm data, showed an unusual ring-like structure
in the total intensity map.

At 6 cm (4.85 GHz), the source has a measured
single-dish
flux density of $300 \pm 38$ mJy (Becker {\it et al.} 1991), and at 20 cm (1.4 GHz)
 $804 \pm 60$ mJy (White \& Becker 1992), indicating an integrated spectral
index  $\alpha \sim -0.8$ ($S_{\nu} \sim \nu^{\alpha}$), which is
moderately steep.
The radio structure in MG0248+0641 (probably 4C+06.13) was first
mapped by Lawrence {\it et al.} (1986) as part of a
program to discover gravitational lenses
in the MIT-Green Bank (MG) catalog. The observations were made with the NRAO
\footnote{The National Radio Astronomy Observatory (NRAO) is a facility of the
National Science Foundation operated under cooperative agreement by Associated
Universities, Inc.}
Very Large Array (VLA) A-array configuration for an on-source
integration time of $\sim$ 2 minutes. Although the resolution
was $\sim 0\farcs3$, the original calibration and mapping of the data were not
good enough to uncover the detailed structure of the extended emission.

As rings are a rare type of morphology within the MG-VLA sample,
MG0248+0641 has now been observed with the VLA at 2 cm (14.9 GHz),
3.6 cm (8.4 GHz) and 6 cm (4.6 GHz), and with
MERLIN at 18 cm (1.6 GHz). Optical and infrared observations of the field
containing the source have been made in R-band with the 2.4 m Hiltner
telescope of the Michigan-Dartmouth-MIT (MDM) Observatory and
in K-band with the KPNO 2.1 m telescope, respectively.
The MDM 2.4 m and MMT have been used to obtain spectra of optical
sources within one arcminute of MG0248+0641.
These observations were carried out initially to determine the lensing
nature of this radio source.

\section{Observations}

New VLA observations of MG0248+0641 have been made in the
A-array at 3.6 and 6 cm in August 1995, and in B-array
at 2 cm in October 1995. The on-source integration times of the 2, 3.6 and 6 cm
observations were 40, 20, and 20 minutes respectively.
Calibration and mapping of these data were
performed using standard AIPS procedures. The maps were further
self-calibrated, yielding final rms noise levels of typically twice
the thermal limit.

The 1995 VLA maps are shown in Fig. 1. The overall structure of the source
is a double-sided core-jet-hotspot system. The 3 bright
peaks correspond to the central core and two  hotspots, with
the core being located at $\alpha$=02:48:58.28,
$\delta$=06:41:43.6 (J2000), with an astrometric accuracy
of $\sim 0\farcs2$ (Lawrence {\it et al.} 1986).
In the 3.6 cm map, the integrated flux densities of the core, western and
eastern hotspots are 9, 23 and 3.2 mJy, respectively.
The noise level
in this map is  $\sim 0.1$ mJy beam$^{-1}$.
The western lobe contains
a prominent ring-like structure in
between the core and western hotspot, and there appears
to be
a large extended area of diffuse emission north of this ring, which
is most distinctively imaged in the 3.6 cm map.
Diffuse
low surface-brightness emission envelopes the ring and the eastern jet,
though this is not well
imaged by the CLEAN deconvolution algorithm. We do find several knots
within this diffuse emission, on both sides of the core.
At the sensitivity of our observations,
there is no evidence for an undisturbed jet to the
west of the core, but we detect a partial
jet east of the core. In Fig. 2 we show the spectral index distribution of
MG0248+0641, based on the VLA 2 and 6 cm observations, where the 6 cm
A-array emission
has been convolved to the 2 cm B-array resolution. The ring-like structure
has a spectral index of $\sim$ -1.0, whereas the
core component is found with a spectral
index of $\sim$ -0.6. The two hotspots are found with a varying spectral index
distribution, with an index
 value close to $\sim$ -0.5 at the ends closest to the
core, and highly steep values $\sim$ -1.5 at the post-shock endpoints.
All of these values are consistent with the integrated index
 $\alpha \sim -0.8$ from
the single-dish measurements.

VLA linear polarization maps were made from all of the 1995 data, at each of
the observed wavelengths, using observations of 3C48 as a polarization
calibrator based on the polarization angle values given in Perley (1982).
In Fig. 3, we show the total intensity contour plots of the radio emission,
onto which we have overlaid the fractional linear polarization electric
field E vectors. The ring exhibits high fractional linear polarization, from
 20\% to 70\%, with
vectors oriented radially outward from the ring center.
The orientations of these vectors do not change by more than $\sim$
10 degrees between 2 and 6 cm, indicating a small
Faraday rotation in this wavelength range.
Therefore, these vectors should be perpendicular to the
magnetic flux density B field lines in the emitting plasma of the
radio galaxy, assuming that the radiation is primarily due to the synchrotron
process.
The hotspots are 12\% polarized.
The diffuse emission north of the ring has a fractional polarization of
only
$\sim$ 5\%, and it is noted that this falls to an unmeasurable value in the
central and brightest part of this emission.
The partial jet to the east of the core is 15\% polarized.
Thus, emission from non-ring portions of this source are weakly polarized
($\leq$ 15\%), while the ring and its associated structures are highly
polarized. In the brighter, outermost part of the eastern jet, the polarization
vectors are oriented along the axis of the jet, but rotate towards
the transverse
direction as the jet arrives at the hotspot.

We also observed MG0248+0641 with the MERLIN array in May 1997,
using 7 antennae at 18 cm,
with an on-source integration time of $\sim$ 8 hours.
The observations of the source were phase referenced to the compact
calibration source 0246+061 (Browne {\it et al.} 1997).
3C286, together with 0246+061, were used as polarization
calibrators, assuming a polarization position angle of 33 degrees for 3C286
at 18 cm.
The data were edited, calibrated and mapped using
standard MERLIN-D programs and procedures within AIPS.
The resulting map (Fig. 4) has a resolution of $0\farcs25$.
In Fig. 4, the core, western and eastern hotspot
components are labeled A, B and C, with integrated
flux densities of 14, 141, and 27 mJy, respectively.
Several resolved knots of emission
in and around the ring are seen.
The partial eastern jet is also resolved into several knots,
and the gap remains between the core and this
jet, at this frequency.
In Fig. 5, we show the MERLIN 18 cm fractional
polarization map, with vectors overlaid onto
the contour map of the total intensity.
At this wavelength, it can be seen that the orientations of the vectors have
changed by more than 70 degrees, especially along the arc-like
feature adjacent to the western hotspot, suggesting a high Faraday rotation
at this longer wavelength.

Based on astrometry using the
Cambridge Automated Plate Measuring machine (APM)
catalog, we found
a faint optical counterpart at $\alpha$=02:48:58.23,
$\delta$=06:41:43.3 (J2000).
This source is $0\farcs6$ southwest of the radio core, but
does not  lie inside the ring.
Bearing in mind an estimated accuracy of 1 arcsecond in the
APM position, we consider this source to be the
optical counterpart of the radio core, and its position
is marked with a cross in Fig. 4.

Optical observations of this field were carried out
in visible wavelengths at the MDM
2.4 m Hiltner telescope in November 1995, and
in the infrared K-band using the InSb $256 \times 256$ array
on the KPNO 2.1 m telescope in January 1996, with total integration times of
50 and 54 minutes, respectively. The data were
reduced using standard procedures in IRAF.
In Fig. 6 we show a 4.5$'$ by 4.5$'$ field
from the 2.4 m telescope, imaged to a limiting magnitude of R $\sim$ 25.5,
with a seeing of 0.89$''$ (FWHM).
The photometric observations of the object considered to be
the optical counterpart of the radio core (marked by a square
box in Fig. 6 and henceforth referred to as the counterpart)
indicate an R magnitude of  $\sim$ 18.6, and  color R-K of $\sim$ 2.8.
This counterpart is
marginally resolved, with a circular isophote of
$\sim 0\farcs92$, and subtracting a stellar point-spread-function
revealed a possible host galaxy underneath.
There are two other nearby sources to the west of the core counterpart
(see Fig. 7). The first is
$3\farcs8$  away
at $\alpha$ = 02:48:57.98, $\delta$ = 06:41:42.9 (J2000)
with R $\sim$ 24.7 (2.1 $\sigma$). The second is $9''$ away
at $\alpha$ = 02:48:57.62, $\delta$ = 06:41:44.3 (J2000),
with R $\sim$ 22.6 (5.9 $\sigma$) and K $\sim$ 17.3 (9 $\sigma$),
and which is slightly extended
with  $\sim 1\farcs06$ (FWHM).
The first source is probably
detected in the infrared image with K $\sim$ 18.8 (3.2 $\sigma$).
However, the positions of the peaks of this
very faint optical and infrared source do not match
precisely; this is understandable as being due to the low signal-to-noise
level in both R and K detections, and also due to known problems with the
KPNO 2.1 m tracking in the infrared image.

Amongst the several objects within 1 arcminute of the counterpart,
there are two galaxies to the south, and two stars,
one to the north and one to the east of it. The star to the
east in obviously a foreground star in our
own galaxy, and therefore has not been observed further.
However, moderate resolution spectra have been obtained in
November 1995 for the three other objects using the MODSPEC
spectrograph on the MDM 2.4 m telescope.
The star to the north is a late type star,
with strong Fe I and Fe II lines, and a thermal spectrum of $T \sim$ 5000 K.
Based on $H\beta$ and [O III] 4959 and 5007 $\AA$ emission
lines, the two galaxies to the south, with
R magnitudes of $\sim$ 15.7 and 16.1, have been shown
to be at the same redshift of 0.1.

In December 1995, we used the Blue Channel
spectrograph on the MMT to obtain a 45 minute spectrum of the core
counterpart.
Two 1200s exposures were combined to produce the final spectrum
in Fig. 8.  The 300 grooves mm$^{-1}$ grating and 1$''$ slit
provided a spectral
resolution of $\sim 5 \AA$ (FWHM). The standard star BD28+4211 was
used for flux calibration. The spectrum was extracted and calibrated
using standard routines in IRAF. The spectrum has an unusually
blue continuum ($F_{\nu} \sim \nu^{\alpha}, \alpha \sim
4.1$), which drops rapidly shortward of 6280 $\AA$, corresponding to
4000 $\AA$ in the rest-frame,
and levels off sharply to a flat continuum at longer wavelengths.
The spectrum shows a broad Mg II 2798 $\AA$ emission line at
 4408 $\AA$,
and two narrow [O III] 4959 $\AA$ and 5007 $\AA$ emission lines, at 7786
$\AA$ and 7881 $\AA$,
respectively. These lines suggest a redshift of 0.57 for the
radio core. The absorption features in the spectrum,
other than prominent atmospheric lines, have equivalent widths no
greater than $\sim$
6 $\AA$, and may result from
stars in the host galaxy of the AGN core counterpart, rather
than from  intervening absorbers.
We have also detected the 3000 $\AA$ bump,
commonly known as the ``weak blue bump'' (Wills {\it et al.} 1985) in QSO
spectra.

\section{Discussion}

Gravitational lensing could produce the ring morphology
observed in MG0248+0641.
As an illustration, we have constructed a simple lens model which accounts
for the major features in the radio ring (Fig. 9).
In this example, an elliptical singular isothermal
potential (Blandford \& Kochanek 1987)
has been placed in front of a background source, which was built up to
reproduce the observed structures.
Probabilistic calculations based on simple lensing optical
depth models (Turner, Ostriker \& Gott 1984)
suggest that there is 90\%
confidence for a lens
to lie between redshifts 0.07 and 0.27. Assuming a point-mass potential
at a redshift of 0.17, which is half the
angular diameter distance to a background source at 772 h$^{-1}$ Mpc
($\Omega_0$ =1, $\Lambda_0$=0, H$_0$ = 70 km s$^{-1}$
 Mpc$^{-1}$ ), the observed ring size can be produced  by a
galaxy with a mass of $\sim$ 1 $\times$ 10$^{11}$ M$_{\odot}$ enclosed
by the projection of the radio ring.
Assuming a factor of five for the ratio of total to enclosed mass,
this would imply a galaxy of mass $\sim$ 5 $\times$ 10$^{11}$ M$_{\odot}$.
The angular size of the ring corresponds
to a line-of-sight velocity dispersion of $\sim$ 235 km s$^{-1}$,
which is typical for galaxies with mass $\sim$ 10$^{11}$ M$_{\odot}$.
The modeled lensing galaxy was given an isophotal ellipticity 0.3,
and oriented to produce the observed gap at the southern edge of the ring.
As shown in Fig. 9, the morphological structure of the western radio lobe
can be easily produced with gravitational lensing.

Lens models cannot explain the observed radio polarization, however.
Polarization is unaffected by lensing, so the radial
orientation of the polarization vectors along the observed ring
would have to correspond to a fortuitous variation along the
proposed background jet. Higher resolution radio observations of
Einstein ring gravitational lens MG1131+0456 have shown
a complex polarization structure (Chen \& Hewitt 1993),
and such complex polarized intensity distributions are likely to
occur in all lenses with ring morphology.
The very high fractional polarization values observed are also
hard to explain in the context of an ordinary background radio jet.
Finally, a very strong objection to the lensing interpretation
is the absence of any detectable lensing galaxy, inside the ring radius.
An elliptical galaxy with one-dimensional velocity
dispersion of 235 km s$^{-1}$ would have a luminosity of $\sim$ 1 $\times$
10$^{11}$ L$_{\odot}$ (Faber \& Jackson 1976), corresponding to an absolute
magnitude M$_R$ $\sim$ -21 (Oegerle \& Hoessel 1991). The {\it k}-corrected
apparent R magnitude is then $\sim$ 18.6. According to Coleman
{\it et al.} (1980), typical apparent R magnitudes
at a redshift of 0.17 are $\sim$ 16.2 for an elliptical, and
$\sim$ 17.5 for a spiral galaxy.
No source has been found in our
optical and infrared
data inside the radio ring, where the galaxy is
expected, down to a background 3 $\sigma$
surface brightness of R $\sim$
24.8 mag arcsec$^{-2}$ and K $\sim$ 19 mag arcsec$^{-2}$.
If MG0248+0641 is indeed lensed, then the optical limiting
magnitudes suggest an
intervening galaxy with mass-to-light ratio at least $\sim$ 250 times
higher than that of a typical galaxy (assumed here to have a mass-to-light
ratio from 5 to 15 $M_{\odot}/L_{\odot}$),
which is very large to be associated with a normal galaxy.
If the MG0248+0641 ring is indeed caused
by gravitational lensing, the implied mass-to-light ratio is estimated to lie
in the range 1000--4000, in solar units. At least three candidate ``dark''
gravitational lenses with similar mass-to-light ratios are known: MG 0023+171
(Hewitt {\it et al.} 1987), MG 2016+112 (Hattori {\it et al.} 1997) and
Q 2345+007 (Duncan 1991).

The marginally-detected optical
source 3.8$''$ west of the radio core is not
likely to be the lens.
This faint optical source (Fig. 7)
lies close to the western hotspot and may well
represent optical emission from this hotspot, or from
a satellite galaxy of the central quasar, assuming it is
at the same redshift as the core.
It is also likely that this source is located beyond a redshift of 0.57,
given that its R-K color is $\sim$ 5.9, but exact determination of the
photometric redshift is impossible due to the
possible presence of dust.

Ring-like morphology in radio sources is also seen in supernova
 remnants, novae,
 planetary nebulae, and H II regions, but the radio and optical properties of
 MG0248+0641 appear to be inconsistent with these interpretations.
Given the integrated spectral index, the ring's angular
size, its radio luminosity,
and the radial linear polarization vectors, these
 possibilities are unrealistic; the ring is too large to be an extragalactic
supernova, nova, planetary nebula, or H II region, and the non-thermal
and steep spectral index distribution of the ring is incompatible with the
thermal sources, which have inverted or flat spectral indices. This rules
out novae, planetary nebulae, and H II regions. Also, if this ring
is due to the interposition of a galactic supernova remnant,
the small size
of the ring suggests that
it is young; between the radio observations of 1980 and
1997, we would have expected to see an expansion in the ring. However,
we have not found any
evidence for a change in the ring's physical size.
Also, the spectral index of the ring, $\alpha$ $\sim$ -1, is too steep
to be a shell-type supernova remnant, where typically a spectral index
of about -0.5 is found. The location of MG0248+0641, far from the
galactic equator at a latitude of $\sim$ $-45^{\circ}$, also suggests
a negligible probability that a galactic
supernova remnant
is superimposed on the  radio lobe.
We searched the literature for
prior radio polarization data on these types of sources, but
found none with radial polarization vectors.

The ring-like structure in MG0248+0641 cannot be easily explained within the
context of standard radio morphologies.
For the observed integrated spectral index of $\alpha \sim$ -0.8, the maximum
degree of linear polarization expected for synchrotron emission
is $\sim$ 72\%, and we see values up to 70\% in the circular structure of
the western lobe.
It is not unusual to see high degrees of fractional linear
polarization, exceeding 50\%, in the lobes of radio galaxies and quasars,
a phenomenon commonly ascribed to shock compression accompanying the transverse
expansion of radio lobes and jets (see the
1988 review by Saikia \& Salter).
Also, very high fractional polarizations, approaching the theoretical maximum,
can only be observed if the angle between the plane of compression and the
line-of-sight is small. These regions are most likely to be found in the
outermost surfaces of the lobes, both where the lobe advances into the
intergalactic medium (IGM), and also where the jet hits the hotspot.
Terminal hotspots with low polarization suggest high inclination angles for
radio galaxies (Laing 1981), i.e. the radio jet axis is pointed close to the
line-of-sight.
In the case of MG0248+0641, the fractional polarizations in the hotspots are
indeed lower than that of the ring-like structure, suggesting that its jets are
in fact inclined at a large angle to the plane of the sky.
Given this suspected inclination in MG0248+0641, the light propagation delay
time between photons from each of the lobes would be expected to cause an
asymmetry in the jet lengths; the observed ratio of jet lengths would then
suggest that it is the (longer) eastern jet which is approaching.
Due to Doppler boosting, this approaching lobe is expected to have a more
visible jet, to be brighter, and more polarized; we find instead that these
particular features are associated with the (shorter) western jet.
Furthermore, if the eastern jet is indeed approaching us, the expected flux
ratio between the western and eastern lobes would be less than unity, which is
again inconsistent with our observations.
Since none of the standard explanations can be used to describe both the flux
and length asymmetries in MG0248+0641, we are led to the conclusion that they
are intrinsic; the western side of this radio galaxy experiences an environment
different from the opposing side.

If the observed ring is due to a disruption of a normal jet
to the west of the core, a severe disturbance is required.
Unusual structures in extragalactic radio sources
have been commonly associated with cD galaxies in cluster centers,
where internal dynamics
may play a key role. The distribution of X-ray surface brightness in clusters
suggests the existence of cooling flows (Fabian {\it et al.}
1991). Many of these cooling flow clusters have a central
dominant galaxy which is associated with a radio source (Burns 1990).
However, in our case we find no evidence for a cluster in
the field of MG0248+0641. Recent ROSAT observations
have not shown any significant X-ray emission, with
an upper limit on the observed flux of $\sim$ 2 $\times$ 10$^{-16}$ W m$^{-2}$
(Wolfgang Brinkmann, personal communication). Assuming a temperature
of 5 kev, this corresponds to a X-ray luminosity of $\sim$ 10$^{37}$ W, at a
redshift of 0.57.
The two galaxies we found to the south of the radio core, at a redshift of
0.1, may well belong to a galaxy group, rather than a foreground galaxy
cluster.

Other sources with prominent circular lobes and
rings have been observed, such as
the southern lobe of 3C310 (van Breugel \& Fomalont 1984), the
northern lobe of 3C219
(Perley {\it et al.} 1980), and Pictor A (Perley {\it et al.} 1997),
with more such sources listed in
van Breugel \& Fomalont (1984).
At a redshift of 0.57, the projected overall size of
MG0248+0641 (45 h$^{-1}$ kpc) and the
average diameter
of the ring (7.5 h$^{-1}$ kpc) are several times smaller than in
these sources. However, the ratio of
ring-to-overall size in MG0248+0641 is comparable to many of these sources.
The rings in 3C219 and 3C310 are similarly polarized with
high fractional values, and
electric field intensity
vectors oriented in the same configuration as in MG0248+0641.
Given that we find no Faraday rotation between 2 and 6 cm,
the polarization vectors in Fig. 3 are parallel to the electric
field, indicating a circumferential
magnetic field.
The eastern side of MG0248+0641 is also compatible with
the southern side of 3C219, which has been modeled by Clarke {\it et al.} 1992.
These similarities have led us to the tentative
conclusion that MG0248+0641 is
a small-scale version of sources such as 3C310 and 3C219.
However, MG0248+0641 is somewhat different from the
known rings in these other sources, which have no
terminal hotspots in their lobes (e.g. 3C310).
This could likely be due to age differences between the respective lobes,
with a smaller, and possibly younger, MG0248+0641 having been energized
relatively recently.

The rings in 3C310 and 3C219 are in fact high-intensity spherical shells,
with no
actual hollow, but rather
a decrease in brightness in their central regions. In MG0248+0641,
we also find that the minimum brightness at the ring
center is slightly higher than the rms noise (off-source), suggesting
that the observed ring-like structure is a projected shell-like feature,
or a ``bubble''. The confinement of such bubbles by a magnetic field and a
hot ambient
medium can easily produce the observed highly linearly polarized emission.
Theoretical predictions of such bubbles, energized by
plasma flows, exist in the literature.
Smith {\it et al.} (1983) have predicted such shells
of hot gas to be blown out by weak jets, through jet choking and other
instabilities in the energy transportation.
In 3C310, an
optical source has been detected next to the optical counterpart of its
core, which has led van Breugel \& Fomalont (1984) to conclude that
the release of energy in the form of bubbles can be triggered by infalling
gas from tidal interaction with this companion. Also, Sadun \& Hayes (1993)
discovered an optical companion to the core counterpart of Hercules A, with a
separation $\sim$ 4$''$, and galaxy pairs are found in
3C219 separated by $\sim$ 8$''$ (Crane, Tyson \& Saslaw 1983).
Recently, Morrison \& Sadun (1996) have argued for a two-stage origin
for the multiple rings seen within some extended radio lobes: in the first
stage, a periodic outflow and weak shocks form the initial shell structure,
with a drift that moves them around,
and then the low-pressure portions of these bubbles are filled up with
energized electrons in the second stage. The periodic tidal forces from the
nearby perturbing companion determine the density modulations, which in turn
produce the series of multiple shells.

The production of a shell in the western lobe
may also be due to a jet instability.
Theoretical predictions suggest that
instabilities are likely to occur near the nuclear regions of
small and
less energetic radio galaxies (Smith {\it et al.} 1983, and
references therein), with
more energetic radio galaxies requiring an external source, such as
those arising in tidal interactions
with other galaxies. The detection of an optical
source $3\farcs8$ away from the core counterpart, at the location near
the western hotspot of MG0248+0641, is promising within the
context of the model of
van Breugel \& Fomalont (1984).
If this is indeed a companion to the central
optical source, a tidal interaction may have produced the required
instability, allowing a recent outflow of energetic plasma
into the western radio jet.
The circular structure may also
have been due to a past, more active phase of the jet, causing
a surge in the gas input to the lobes, and hence a more symmetrical
subsequent expansion.
Finally, it is also likely that the asymmetry in MG0248+0641
is due to the disruption of the western radio jet as it runs into
its companion galaxy.

Since the current models cannot fully describe the energetics required to
produce the observed high-intensity shell features, more investigative
theoretical work is required in order to describe the above processes, and
possibly others not suggested here. The MG0248+0641 field may also be a good
candidate for deep observations with the Keck or HST; it would be interesting
to search for more
direct evidence of galaxy interactions. Also, direct redshift determination of
the optical companion may help in accepting or rejecting the interaction
hypothesis presented here; however, spectroscopic measurements will be
challenging, given its faint magnitude.

\acknowledgments

We would like to acknowledge Rodney Davies for granting us Director's
discretionary time to use MERLIN, Paul Schechter for useful
discussions, Rick Perley for pointing out a problem with our initial
polarization calibration, and Wolfgang Brinkmann for providing us
with results from ROSAT observations of the MG0248+0641 field.
The National Radio Astronomy Observatory is a facility of the National Science
Foundation operated under cooperative agreement by Associated Universities,
Inc.
MERLIN is a national facility operated by the University of Manchester on
behalf of the Particle Physics and Astronomy Research Council.
IRAF is distributed by the National Optical Astronomical Observatories, which
are operated by the Association of Universities for Research in Astronomy,
Inc., under cooperative agreement with the National Science Foundation.
The MDM Observatory is operated by a consortium of the University of Michigan,
Dartmouth College and the Massachusetts Institute of Technology.
The Multiple Mirror Telescope (MMT) is operated as a joint facility of
the Smithsonian Institution and the University of Arizona by the Multiple
Mirror Telescope Observatory, and is located on the grounds of the Fred
Lawrence Whipple Observatory of the Smithsonian Astrophysical Observatory on
Mount Hopkins.
This research was supported by  NSF grant AST92-24191 at MIT.
JL gratefully acknowledges support from NSF grant AST93-03527.

\clearpage

\figcaption[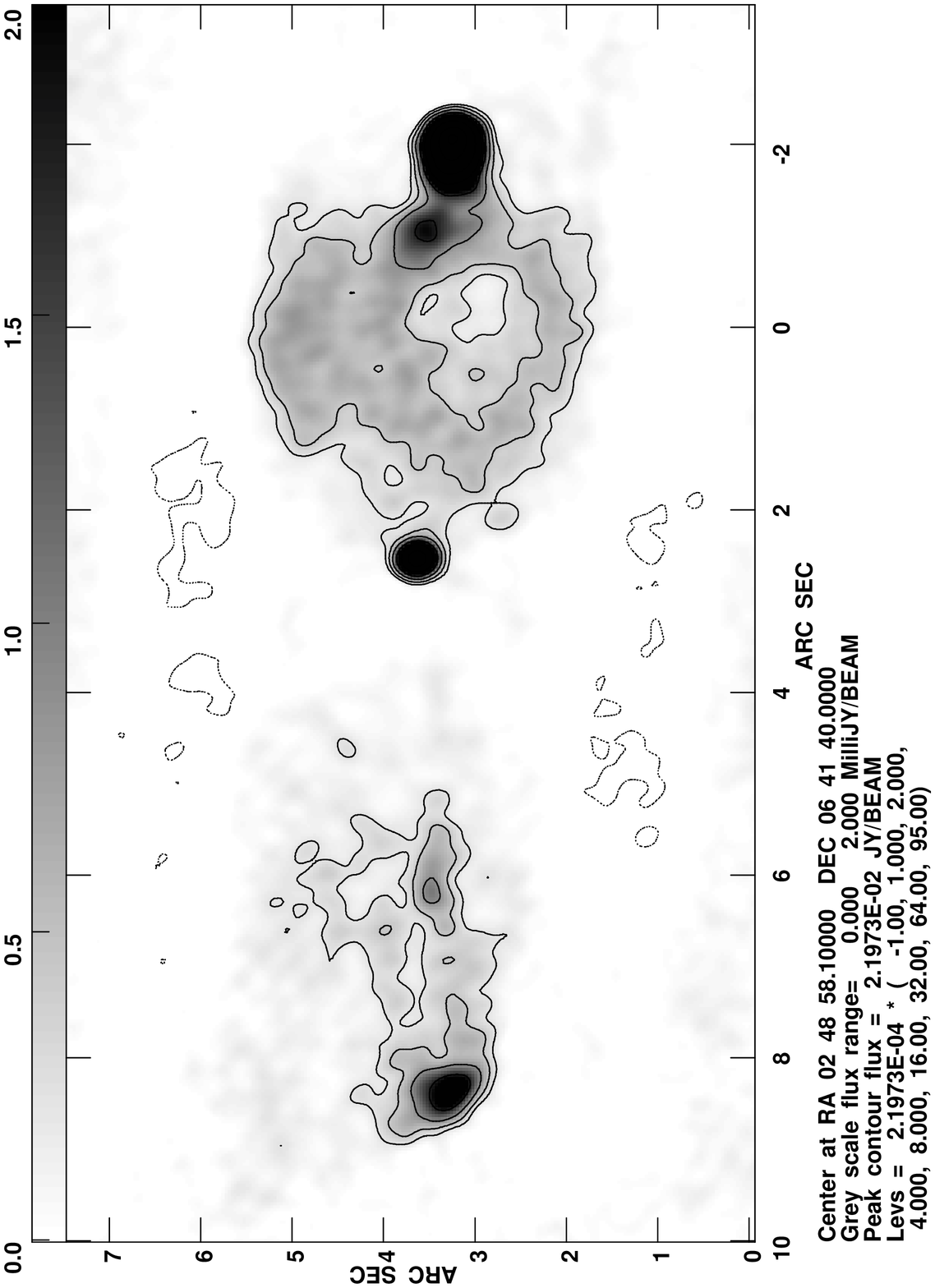]{VLA A-array 3.6 cm
total intensity maps of MG0248+0641. The beam size in the 3.6 cm map is
$0\farcs3$, and the peak total
intensity is 22 mJy beam$^{-1}$. \label{fig1}}

\figcaption[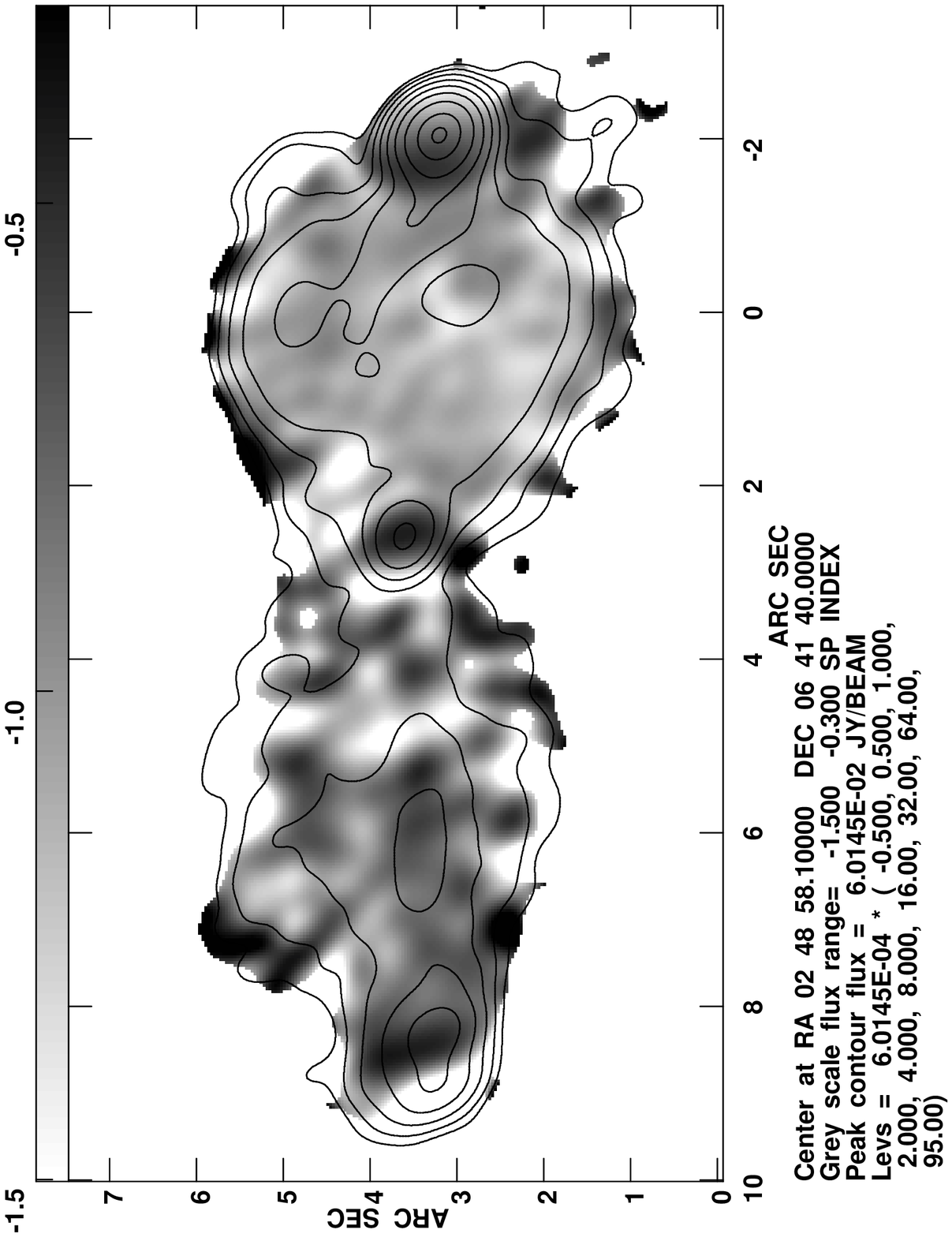]{Spectral index distribution of MG0248+0641,
based on VLA 2 and 6 cm observations (grey scale).
The contours represent A-array 6 cm total intensity emission, but
convolved down to the B-array 2 cm resolution.
The ring-like structure has a spectral index of
$\alpha \sim -1.0$, whereas the unresolved core and hotspot
components have flat spectral indices $\sim$ -0.5. \label{fig2}}

\figcaption[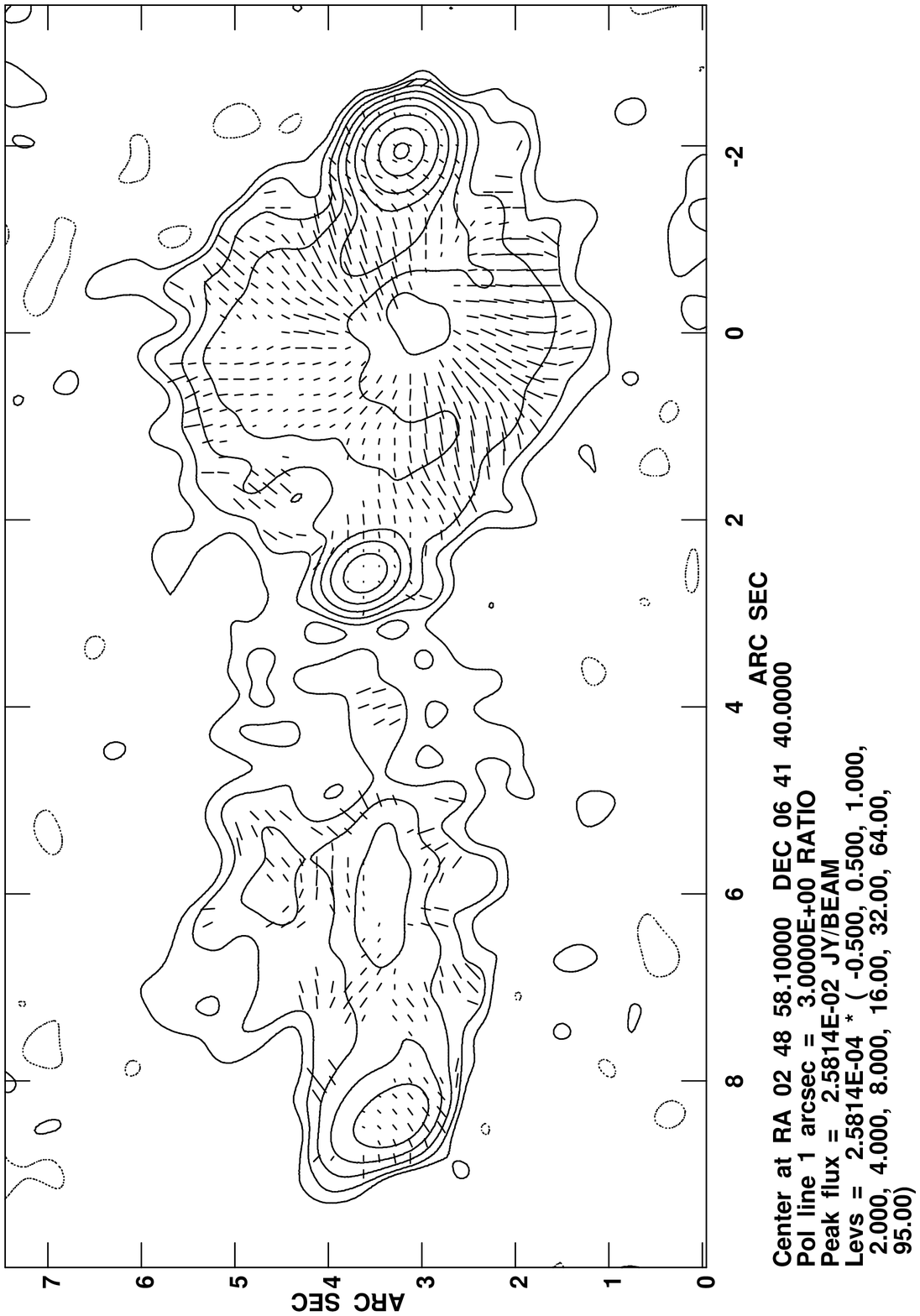]{VLA A-array 6 cm
total intensity contour maps of
MG0248+0641, onto which linear
polarization vectors have been overlaid. The natural
weighted beam size in the 6 cm map is $0\farcs48$,
and is shown in the bottom left corner. In all 3 wavelengths,
the fractional polarization vectors are scaled such that $0\farcs33$
corresponds to 100\%, and they
are oriented along the electric field.\label{fig3}}

\figcaption[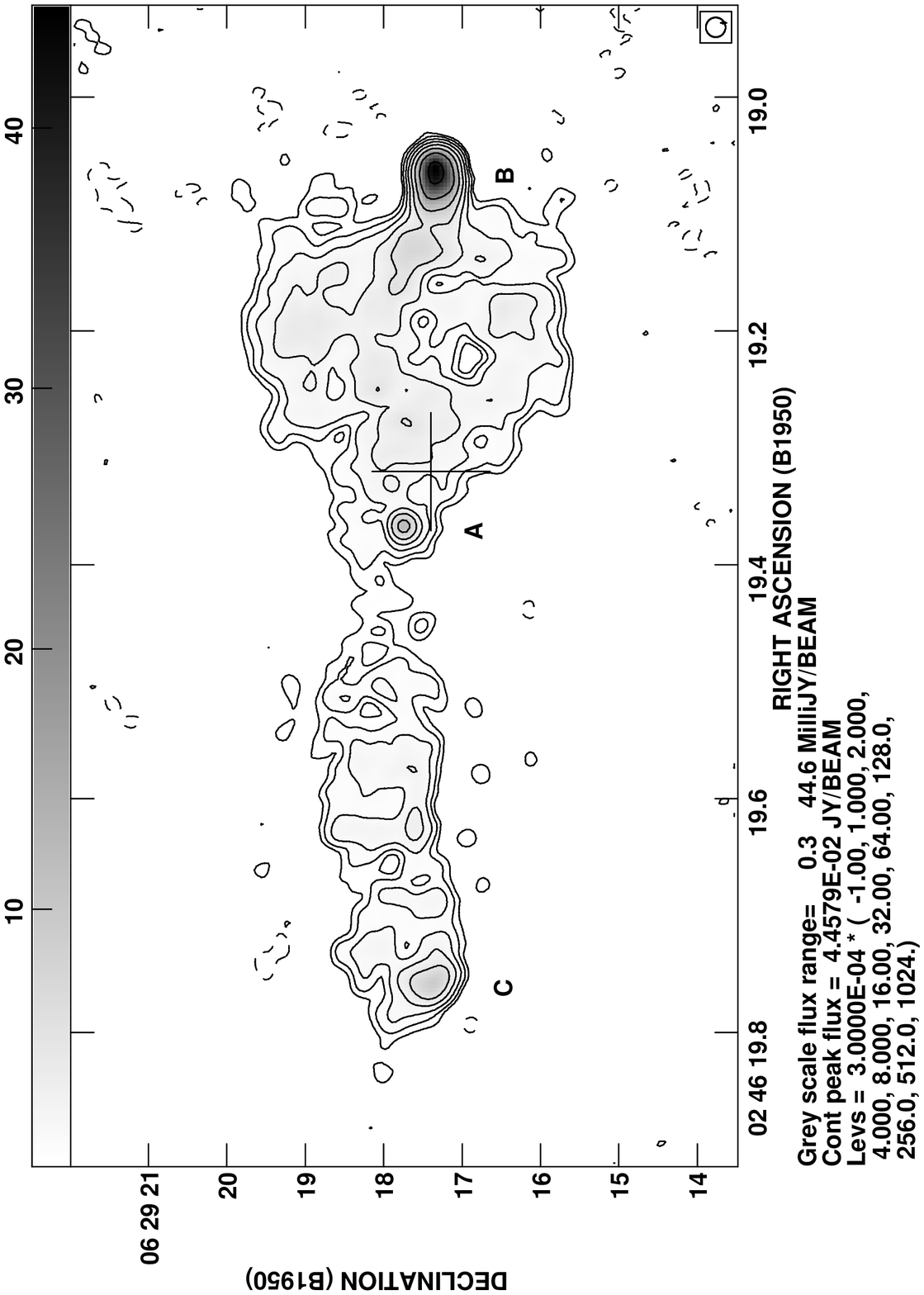]{MERLIN 18 cm total intensity
contour map of MG0248+0641, with a restored
beam size of $0\farcs25$, which is shown in the bottom right corner.
The optical counterpart position from the APM catalog is marked with
a cross, scaled to the $\pm$ $1\farcs0$ APM astrometric uncertainty.
\label{fig4}}

\figcaption[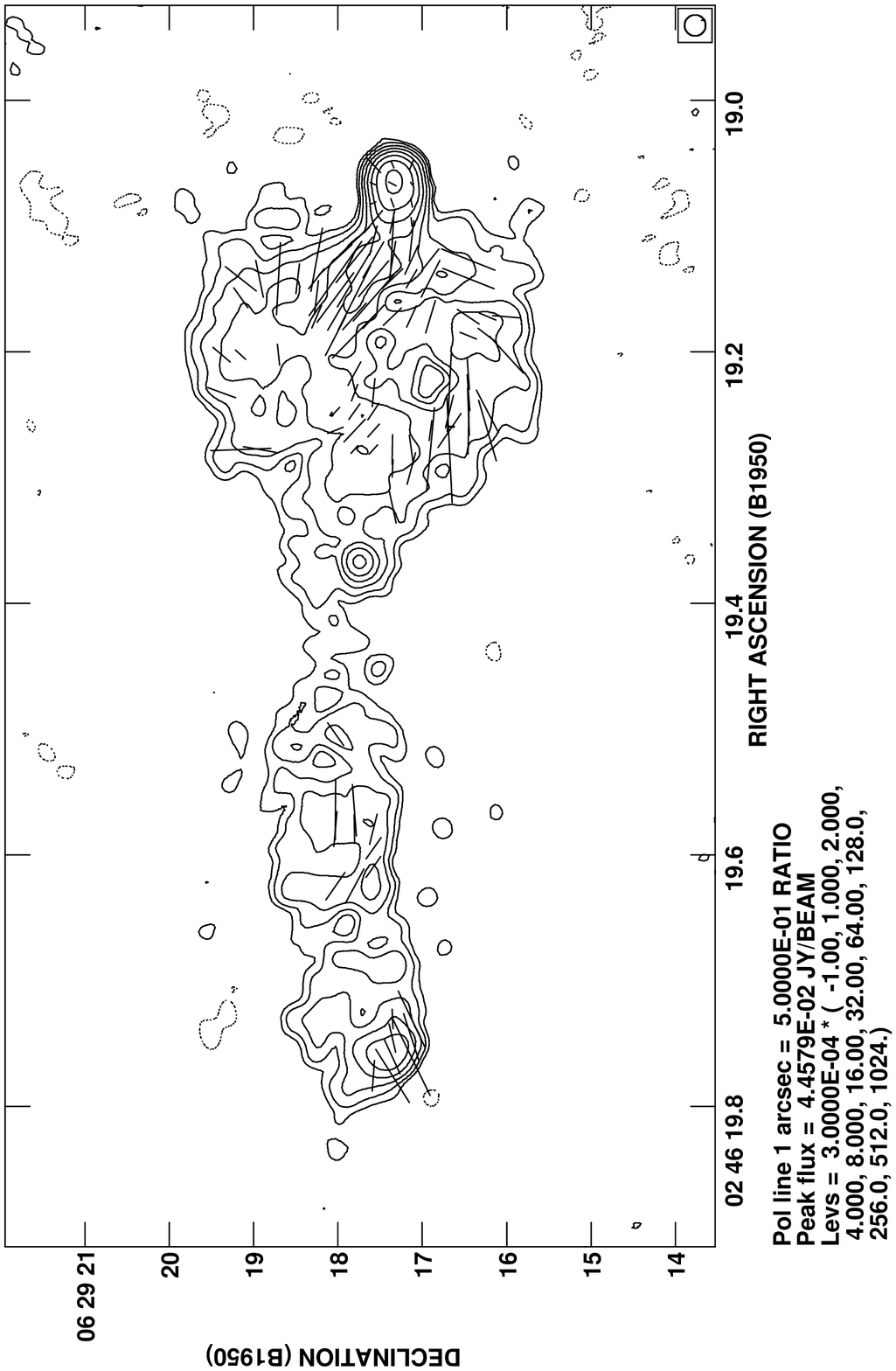]{MERLIN 18 cm total intensity
contour map of MG0248+0641, onto which linear
fractional polarization vectors have
been overlaid. The vectors are
scaled such that one arcsecond of vector length
corresponds to 50\%, and they are oriented along the electric
field. \label{fig5}}

\figcaption[fig6.gif]{R band optical CCD image, with a
total integration time
 3000 seconds, of the MG0248+0641 field ($4.5'$ by $4.5'$),
observed with the MDM 2.4 m telescope.
North is left; east is down. The
counterpart is marked with a square box. The limiting
magnitude is $\sim$ 25. The pixel scale is $0\farcs275$. \label{fig6}}

\figcaption[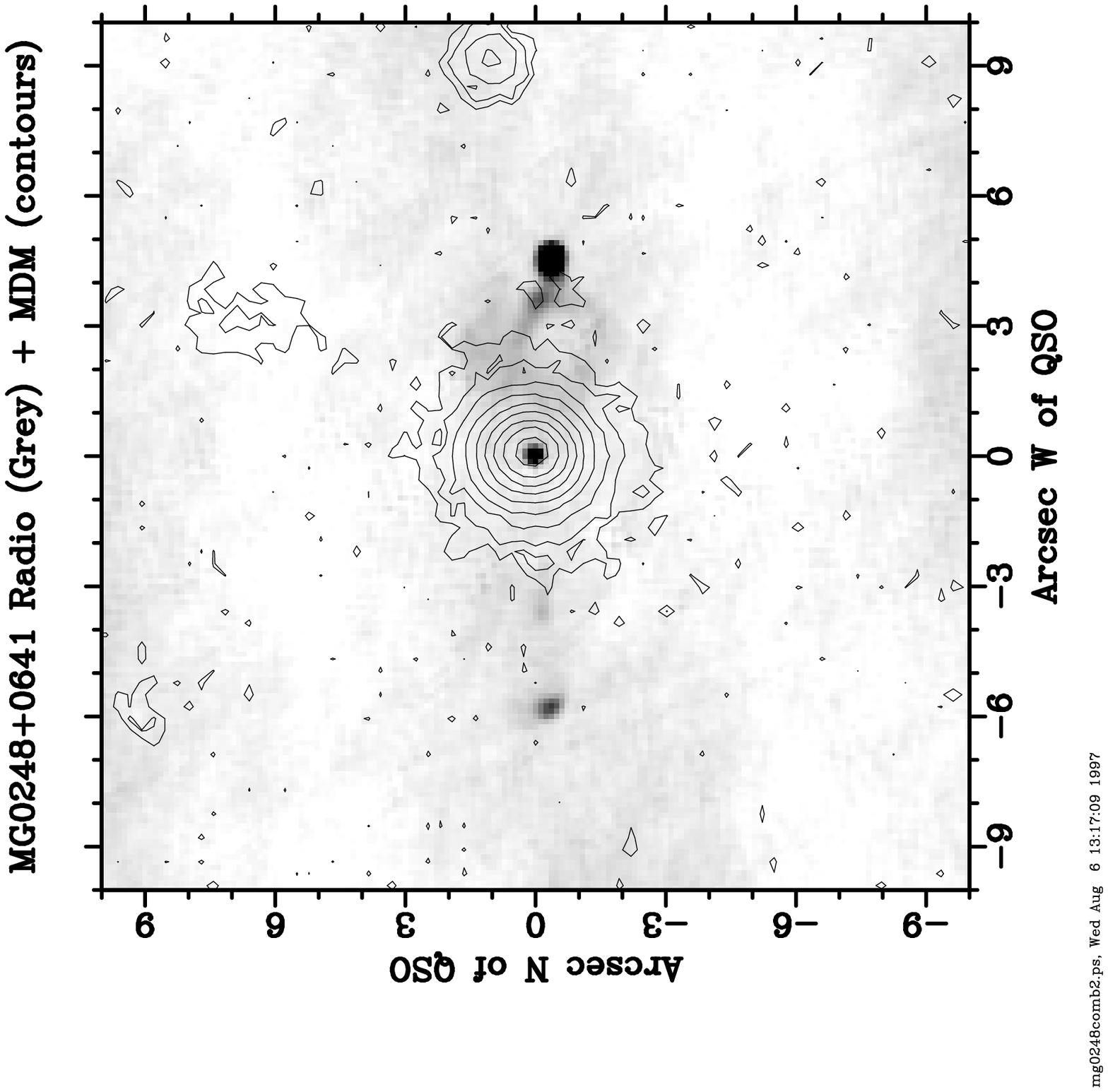]{MDM 2.4 m R band
image overlaid on the 3.6 cm VLA map, with
contours in steps of the 2 $\sigma$ noise level in the optical data.
We have positioned the optical core counterpart such that it lies
on top of the radio
core. A faint optical source is marginally detected
 close to the western hotspot, at a level of 2.1 $\sigma$.
\label{fig9}}

\figcaption[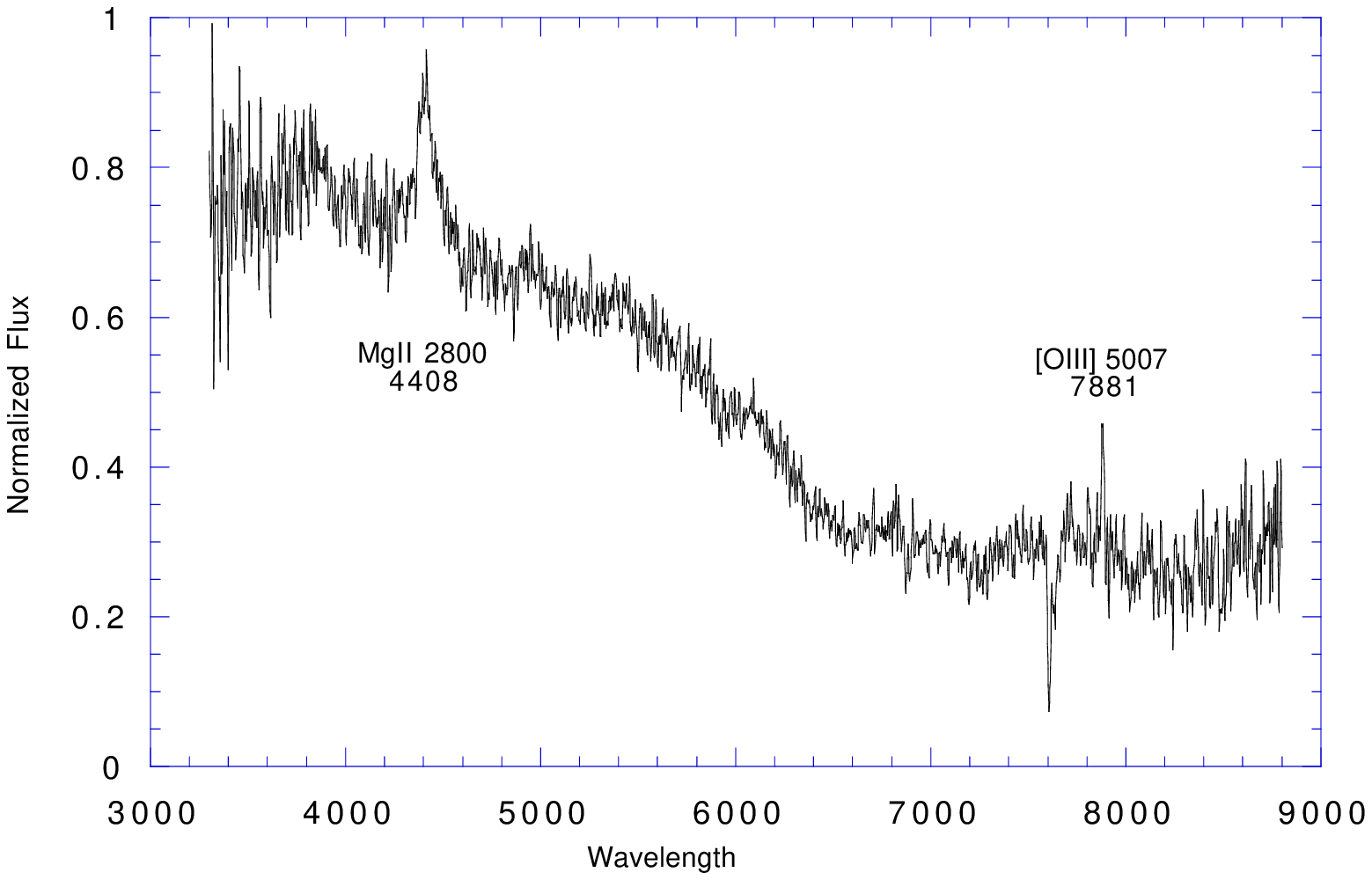]{The optical spectrum of the counterpart of
core component (A), taken at
the MMT with the Blue Channel Spectrograph.
Wavelength is in angstroms, and the flux is normalized.
The spectrum shows the ``weak blue bump'',
and Mg II and [O III] emission lines, from which we
calculate a redshift of 0.57. The prominent absorption
near 7600 $\AA$ is a sky feature. \label{fig7}}

\figcaption[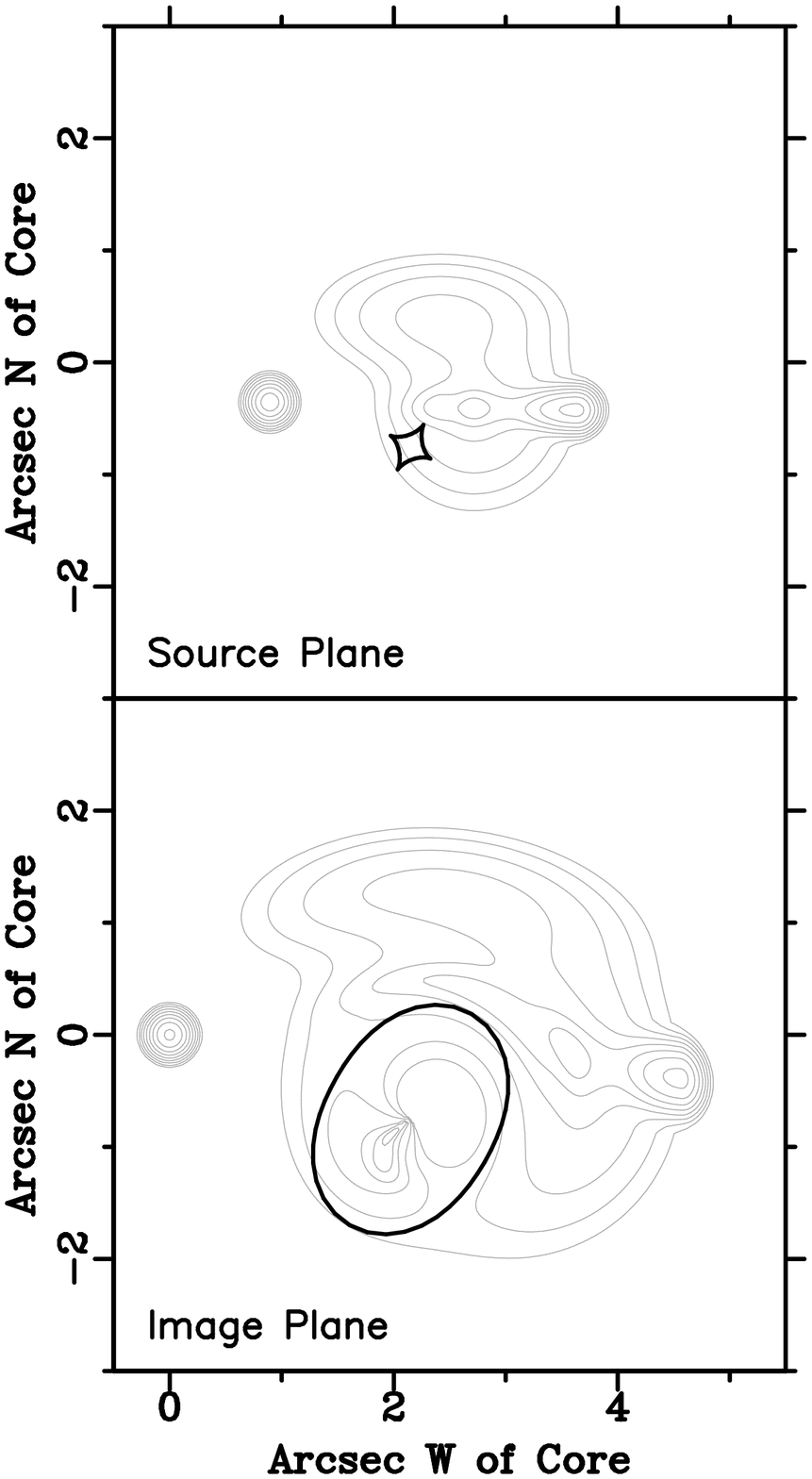]{Lens model interpretation for MG0248+0641.
The contours in the source plane show the background source
as it would appear without lensing,
and the diamond caustic is drawn in bold.
The image plane shows the same source projected through the
lens model, with the tangential critical line drawn in bold.
The source was constructed to roughly produce the observed structures.
\label{fig8} }

\begin{figure}
\plotone{fig1b.ps}
\end{figure}
 
\begin{figure}
\plotone{fig2.ps}
\end{figure}
 
\begin{figure}
\plotone{fig3a.ps}
\end{figure}
 
\begin{figure}
\plotone{fig4.ps}
\end{figure}
 
\begin{figure}
\plotone{fig5.ps}
\end{figure}
 
\begin{figure}
\plotone{fig7.ps}
\end{figure}

\begin{figure}
\plotone{fig8.ps}
\end{figure}
 
\begin{figure}
\plotone{fig9.ps}
\end{figure}

\end{document}